**Quasi-Normal Mode Theory for Thermal Radiation from Lossy and Dispersive Optical Resonators**

Baoan Liu, Jiayu Li, and Sheng Shen*

Department of Mechanical Engineering, Carnegie Mellon University, Pittsburgh, PA, 15213, USA

*Corresponding author. E-mail: sshen1@cmu.edu

**Abstract**

Although optical resonators are widely used for controlling and engineering thermal radiation, what has been lacking is a general theoretical framework to elucidate the thermal emission of optical resonators, and guide the design and application of thermally driven optical resonators. We developed a general and self-consistent formalism to describe the thermal radiation from arbitrary optical resonators made by lossy and dispersive materials like metals, with the only assumption that the resonators have a single predominant resonance mode. Our formalism demonstrates that the thermal emission of an optical resonator is maximized when the mode losses to the emitter and the absorber (or far-field background) are matched, and meanwhile the predominant resonant modes are electrically quasi-static. By efficiently calculating the resonant modes using the finite element methods, our formalism serves as a general principle of designing the arbitrary optical resonator thermal emitters with perfect or maximized emission.

A thermal light source like a blackbody or the incandescent filament of a light bulb usually has a broad emission spectrum. However, in many energy related applications such as infrared sensing [1], thermophotovoltatics [2,3], radiation cooling [4] and thermal circuits [5,6], thermal emission is in general required to be much narrower than that of a blackbody. A common paradigm for realizing narrow band thermal radiation is to utilize optical resonators including optical antennas [7], photonic crystal cavities [8], and graphene nanostructures [6,9]. According to the Purcell effect [10], thermal radiation from an optical resonator can be dramatically modulated by the resonance mode designed in the infrared range, leading to the narrow band thermal emission peaking at the resonant frequency. Since thermal radiation is intrinsically weaker than the infrared light from the light sources driven by electrical power, e.g. laser or LED, it is critical but challenging to maximize the emission power of a narrow band thermal emitter. To reveal the principle of maximizing the thermal radiation from an optical resonator, a semi-analytical formalism based on coupled mode theory has recently been proposed to phenomenologically model an optical resonator thermal emitter as a general resonant system with different energy loss mechanisms [5,8,11,12]. It was discovered that the peak of radiation intensity in an emitter reaches a maximum when the energy loss rate to an absorber (or far-field background) is equal to the one to the emitter. Despite the success of couple mode theory in understanding the thermal radiation from optical resonators, this formalism is not directly consistent with electromagnetic wave theory, and therefore the energy loss rate lacks a clear definition from fundamental electrodynamics, which particularly imposes a significant difficulty for calculating the energy loss rates in lossy and dispersive media.

In this Letter, rather than using the phenomenological approach, we develop a general and self-consistent formalism from fluctuational electrodynamics [13] and Quasi-Normal Mode (QNM) theory [14–16] to describe the thermal radiation from optical resonators with arbitrary geometries that are made by lossy and dispersive materials like metals. By only assuming an optical resonator thermal emitter with a single predominant resonance mode, our formalism provides a rigorous definition to the *mode loss* with a closed-form expression by considering the non-Hermitian nature of the lossy resonant mode and the material dispersion. The new formulism also shows that to maximize the narrow band thermal radiation from an optical resonator, not only the mode losses to the emitter and the absorber (or far-field background) require to be matched, but the resonant mode needs to be electrically quasi-static, i.e. the electric field of the resonant mode oscillates in phase. This is intrinsically different from coupled-mode theory. By efficiently evaluating the lossy resonant modes of an optical resonator using finite element methods, the new formalism thus paves the way for designing arbitrary optical resonator thermal emitters with perfect or maximized emission.

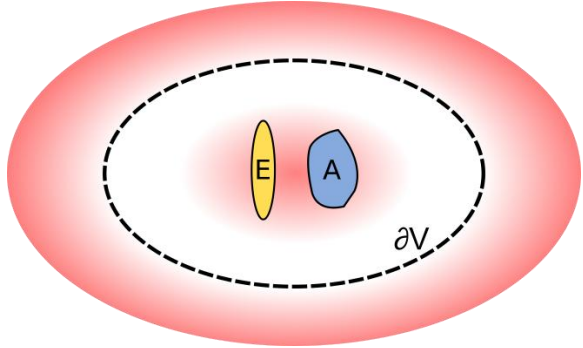

**FIG. 1: (color online.) Schematic of a thermal emitter and a near-field absorber placed in the vacuum background.**

Consider a thermal emitter $V_E$ at the temperature of $T_E$ and a closely separated object $V_A$, where $V_E$ and $V_A$ are placed in vacuum, as shown in Fig. 1. Here, we assume that the materials are non-magnetic, and have isotropic electrical responses. Since the thermal radiation from $V_E$ is physically the emission of electromagnetic waves $[\mathbf{E}(r,\omega), \mathbf{H}(r,\omega)]$ generated by the thermally induced random currents $\mathbf{j}(r,\omega,T_E)$ inside $V_E$, the spectral thermal energy transfer from $V_E$ to $V_A$ is therefore equal to the averaged electromagnetic absorption power of $V_A$,

$$\phi_A(\omega) = \int_{V_A} dr^3 \left\langle \frac{1}{2}\sigma_A \mathbf{E}(r,\omega)^\dagger \cdot \mathbf{E}(r,\omega) \right\rangle, \tag{1}$$

where $\dagger$ denotes the conjugate-transpose, both the field and the current are expressed as 3-by-1 column vectors, and $\sigma$ indicates the electric conductivity of a material, which relates to its permittivity as $\sigma(\omega) = \omega \text{Im}[\epsilon(\omega)]$. Similarly, the spectral energy transfer from $V_E$ to the far-field, $\phi_\infty(\omega)$, equals the integration of the averaged Poynting vector over an enclosure surface $\partial V$, which can be further expressed by the total power from the current sources minus the near-field power absorption in both $V_E$ and $V_A$ according to energy conservation [17],

$$\begin{aligned}\phi_\infty(\omega) &= \int_{\partial V} dr^2 \hat{\mathbf{n}} \cdot \left\langle \frac{1}{2}\text{Re}\left[\mathbf{E}\times\mathbf{H}^\dagger\right]\right\rangle \\ &= -\int_{V_E} dr^3 \left\langle \frac{1}{2}\text{Re}\left[\mathbf{j}^\dagger \cdot \mathbf{E}\right]\right\rangle - \int_{V_E} dr^3 \left\langle \frac{1}{2}\sigma_E |\mathbf{E}|^2 \right\rangle - \int_{V_A} dr^3 \left\langle \frac{1}{2}\sigma_A |\mathbf{E}|^2 \right\rangle,\end{aligned} \tag{2}$$

where $|\mathbf{E}|^2 = \mathbf{E}^\dagger \cdot \mathbf{E}$. Note that we ignore the backward thermal radiation from both $V_A$ and the background to $V_E$. In addition, the spectral energy transfer $\phi(\omega)$ relates to the total power $\Phi$ as $\Phi = \int_0^\infty d\omega \phi(\omega)$.

The spectral energy transfer $\phi_A(\omega)$ and $\phi_\infty(\omega)$ in Eqs. (1) and (2) can be formulated as deterministic expressions, because (i) the electric field emitted by the random currents can be represented as $\mathbf{E}(r,\omega) = i\omega\mu_0 \int_{V_E} dr'^3 \overline{\mathbf{G}}_{\omega,r,r'} \cdot \mathbf{j}(r',\omega,T_E)$, where $\overline{\mathbf{G}}_{\omega,r,r'}$ is the Dyadic Green's function defined as the impulse response of the wave equation $[\nabla\times\nabla\times +\omega^2\mu_0\epsilon(\omega,r)]\overline{\mathbf{G}}_{\omega,r,r'} = \overline{\mathbf{I}}\delta(r-r')$ and $\overline{\mathbf{I}}$ is the 3-by-3 unit matrix [10]; (ii) the autocorrelation of the random currents $\mathbf{j}(r,\omega,T_E)$ is characterized by the fluctuation-dissipation theorem [13] as

$$\left\langle \mathbf{j}(r,\omega,T_E)\mathbf{j}^\dagger(r',\omega,T_E)\right\rangle = \frac{4}{\pi}\sigma_E \Theta(\omega,T_E)\delta(r-r')\overline{\mathbf{I}}, \tag{3}$$

where $\Theta(\omega, T) = \hbar\omega/[\exp(\hbar\omega/k_B T) - 1]$ is the Planck distribution. Substituting the Dyadic Green's function and Eq. (3) into Eq. (1), $\phi_A(\omega)$ becomes

$$\phi_A(\omega) = \frac{\Theta}{2\pi} 4\omega^2 \mu_0^2 \text{Tr}\left[\int_{V_A} dr^3 \int_{V_E} dr'^3 \, \sigma_A \sigma_E \bar{\mathbf{G}}^\dagger_{\omega,r,r'} \cdot \bar{\mathbf{G}}_{\omega,r,r'}\right]. \quad (4)$$

where Tr[·] denotes the trace of the matrix. Similarly, for the first term on the right-hand side of Eq. (2), we have $-\int_{V_E} dr^3 \frac{1}{2}\text{Re}[\mathbf{j}^\dagger \cdot \mathbf{E}] = \frac{\Theta}{2\pi} 4\omega\mu_0 \, \text{Tr}\left[\int_{V_E} dr^3 \sigma_E \text{Im}[\bar{\mathbf{G}}_{\omega,r,r}]\right]$. Together with Eq. (4), $\phi_\infty(\omega)$ becomes

$$\phi_\infty(\omega) = \frac{\Theta}{2\pi}\left\{4\omega\mu_0 \text{Tr}\left[\int_{V_E} dr^3 \sigma_E \, \text{Im}\left[\bar{\mathbf{G}}_{\omega,r,r}\right]\right] - 4\omega^2 \mu_0^2 \text{Tr}\left[\int_{V_E} dr^3 \int_{V_E} dr'^3 \, \sigma_E^2 \bar{\mathbf{G}}^\dagger_{\omega,r,r'} \cdot \bar{\mathbf{G}}_{\omega,r,r'}\right]\right.$$
$$\left. -4\omega^2 \mu_0^2 \, \text{Tr}\left[\int_{V_A} dr^3 \int_{V_E} dr'^3 \, \sigma_A \sigma_E \bar{\mathbf{G}}^\dagger_{\omega,r,r'} \cdot \bar{\mathbf{G}}_{\omega,r,r'}\right]\right\}. \quad (5)$$

Eqs. (4) and (5) are the general expressions for near-field and far-field thermal radiation, respectively. The only assumption is the quasi-thermal equilibrium in $V_E$, i.e. uniform temperature inside the emitter $V_E$.

If the thermal emitter in Fig. 1 is simultaneously an optical resonator, its far-field and near-field thermal radiation can be narrow-band and modulated by a predominant resonant mode. In this scenario, we can expand the Dyadic Green's function $\bar{\mathbf{G}}_{\omega,r,r'}$ in terms of the resonant modes based on QNM theory, especially for dispersive and lossy materials. For an optical resonator in vacuum (as shown in Fig. 1), its resonant modes are naturally defined as the eigen-solutions of the source-free Maxwell equations

$$\begin{aligned}\nabla \times \mathbf{E_n}(r) &= i\omega_n \mu_0 \mathbf{H_n}(r) \\ \nabla \times \mathbf{H_n}(r) &= -i\omega_n \epsilon(\omega_n, r) \mathbf{E_n}(r)\end{aligned} \quad (6)$$

Here, the electromagnetic fields $[\mathbf{E_n}(r), \mathbf{H_n}(r)]$ satisfy the outgoing wave boundary condition at $|r| \to \infty$ [14,15]. In Eq. (6), $\omega_n$ is the eigen-frequency, which is a complex number in the cases that the resonant mode is lossy. Specifically, $\text{Re}[\omega_n]$ equals the resonant frequency, and $\text{Im}[\omega_n]$ indicates the mode loss rate. Note that a thermal emitter has $\text{Im}[\omega_n] < 0$ because it must contain dissipative materials to radiate, according to Eq. (3).

Since $\bar{\mathbf{G}}_{\omega,\mathbf{r},\mathbf{r'}}$ is essentially the impulse response of the Maxwell equations, it can be mathematically expanded in terms of the eigen-solutions of the Maxwell equations, when the resonant modes are orthonormal and complete [18] as

$$\bar{\mathbf{G}}_{\omega,\mathbf{r},\mathbf{r'}} = \sum_n \frac{\mathbf{E_n}(r) \cdot \mathbf{E_n^\dagger}(r')}{\omega \mu_0 (\omega_n - \omega) N_{nn}}, \tag{7}$$

where $N_{nn}$ is the orthonormal factor for the mode $n$ to itself. However, the orthonormality and completeness of the lossy resonant modes are difficult to be defined and justified. Until recently this difficulty is resolved by QNM theory [14,15]. Since we are only interested in the frequencies in the vicinity of the resonant frequency of a predominated resonant mode, QNM theory proves that the orthonormality and completeness in this condition are approximately held by defining the orthonormal factor $N_{nm}$ [15,19] as

$$N_{nm} = \int_{V_\infty} dr^3 \left[ \frac{\partial \omega \epsilon(r,\omega)}{\partial \omega} \mathbf{E_n}^T \cdot \mathbf{E_m} - \frac{\partial \omega \mu}{\partial \omega} \mathbf{H_n}^T \cdot \mathbf{H_m} \right]\bigg|_{\omega=\omega_n}, \tag{8}$$

where $X^T$ denotes the matrix transpose and $V_\infty$ indicates the entire space. The "quasi-" completeness requires that (i) only the field inside or in proximity to the optical resonator can be expanded by the lossy resonant modes [14,15]; (ii) the lossy resonant modes used in the expansion account for all the important energy decay channels [15]. Both of these two requirements are satisfied in our cases, because (i) Eqs. (4) and (5) only evaluate $\bar{\mathbf{G}}_{\omega,\mathbf{r},\mathbf{r'}}$ with $r, r' \in V_E \cup V_A$; (ii) the thermal emitters studied in our cases are assumed to have a predominant resonant mode. Furthermore, the "quasi-" orthonormality of the lossy resonant modes is approximately hold for $\omega \approx \omega_n$. The QNM theory for expanding the field by using the lossy resonant modes has recently attracted massive attentions [14]. Several reports have demonstrated the good accuracy of QNM theory by comparing the directly simulated field profile $[\mathbf{E}(r,\omega), \mathbf{H}(r,\omega)]$ near the resonant structures emitted by a dipole source with the expansion of its lossy resonant modes, and good agreements are observed [15,20,21]. As a result, QNM theory justifies that Eq. (7) with the definition of $N_{nn}$ in Eq. (8) is approximately held for the lossy resonant modes expansion near the resonant frequency $\omega \approx \text{Re}[\omega_n]$, which can then be substituted into Eqs. (4) and (5).

To clarify the effect of an individual resonant mode on thermal radiation, we simplify the physics by assuming the non-degeneracy of the resonant modes, and their resonant frequencies $\text{Re}[\omega_n]$ are highly distinct from each other. Consider the frequencies around the resonant frequency of the predominant resonant mode, i.e. $\omega \approx \text{Re}[\omega_n]$. The near-field radiative energy transfer from $V_E$ to $V_A$ in Eq. (4) becomes

$$\phi_A(\omega) = \frac{\Theta(\omega,T)}{2\pi}\left[\frac{\text{Re}[\omega_n]^2}{\text{Re}[\omega_n]^2 + 4Q_n^2(\text{Re}[\omega_n]-\omega)^2}\right]\psi_A(\omega)$$
$$= \frac{\Theta(\omega,T)}{2\pi}L(\omega)\psi_A(\omega) \quad (9)$$

where $L(\omega)$ is the Lorentzian line shape function with the peak at the resonant frequency $\omega = \text{Re}[\omega_n]$, and $Q_n = \left|\frac{\text{Re}[\omega_n]}{2\text{Im}[\omega_n]}\right|$ is the Q-factor of the resonant mode $n$. Eq. (9) predicts the line shape of the peak in thermal radiation spectrum, where the peak width is inversely proportional to the quality factor $Q_n$, and the peak height is determined by $\psi_A(\omega)$ that can be expressed as

$$\psi_A(\omega) = \frac{16}{\text{Im}[\omega_n]^2}\frac{1}{|N_{nn}|^2}\left[\int_{V_A}dr^3\frac{1}{2}\sigma_A(\omega)|\mathbf{E_n}(r)|^2\right]\left[\int_{V_E}dr'^3\frac{1}{2}\sigma_E(\omega)|\mathbf{E_n}(r')|^2\right]. \quad (10)$$

$\phi_A(\omega)$ does not exactly follow the Lorentzian line shape when the materials are dispersive. Nevertheless, for the cases that $\sigma_A$ and $\sigma_E$ do not abruptly vary at $\omega \approx \text{Re}[\omega_n]$, $\psi_A(\omega) \approx \psi_A(\text{Re}[\omega_n])$ indicates the peak of the near-field energy transfer power density. Eq. (10) alone fails to describe the dominant mechanism for maximizing the peak value $\psi_A(\omega)$ and thus guide the emitter design. Here, we further express $\text{Im}[\omega_n]$ using the resonant modes $[\mathbf{E_n}(r), \mathbf{H_n}(r)]$. From Eq. (6), it has the mathematical identity $\int_{\partial V} dr^2 [\mathbf{E_n} \times \mathbf{H_n^\dagger} + \mathbf{E_n^\dagger} \times \mathbf{H_n}] = -2 \int_V dr^3 (|\mathbf{H_n}|^2 \text{Im}[\omega_n] \mu_0 + |\mathbf{E_n}|^2 \text{Im}[\omega_n \epsilon(\omega_n)])$, where $V$ is the volume enclosed by $\partial V$ (as shown in Fig. 1), and $\text{Im}[\omega_n \epsilon(\omega_n)] = \sigma(\text{Re}[\omega_n]) + \text{Im}[\omega_n] \text{Re}\left[\frac{\partial \omega \epsilon}{\partial \omega}\Big|_{\omega=\text{Re}[\omega_n]}\right] + o(\text{Im}[\omega_n]^2)$. As a result, $\text{Im}[\omega_n]$ can be expressed as

$$\frac{1}{-\text{Im}[\omega_n]} \approx \frac{\frac{1}{2}\int_V dr^3 \left[\text{Re}\left(\frac{\partial \omega \epsilon(r,\omega)}{\partial \omega}\right)_{\text{Re}[\omega_n]} |\mathbf{E_n}|^2 + \mu_0 |\mathbf{H_n}|^2\right]}{\int_{\partial V} dr^2 \frac{1}{2}\text{Re}\left[\mathbf{E_n} \times \mathbf{H_n^\dagger}\right] + \int_{V_E+V_A} dr^3 \frac{1}{2}\sigma(\text{Re}[\omega_n])|\mathbf{E_n}|^2}. \quad (11)$$

Eq. (11) agrees with the conventional definition of the mode loss rate $\tau = -1/\text{Im}[\omega_n]$, which equals the energy stored in the resonator divided by the energy loss per cycle [17]. In addition, the numerator on the RHS of Eq. (11) agrees with the universal description of the energy density, especially in the dispersive materials like metal [22]. Substitute Eq. (11) into Eq. (10), the peak value of the near-field spectral energy transfer from $V_E$ to $V_A$ equals

$$\Psi_A = 4\left(\frac{D_E}{D_E + D_A + D_\infty} F\right)\left(\frac{D_A}{D_E + D_A + D_\infty} F\right), \quad (12)$$

where $D_E = \int_{V_E} dr^3 \frac{1}{2}\sigma_E(\text{Re}[\omega_n])|\mathbf{E_n}(r)|^2$ and $D_A = \int_{V_A} dr^3 \frac{1}{2}\sigma_A(\text{Re}[\omega_n])|\mathbf{E_n}(r)|^2$ represent the mode energy losses due to the resistive dissipation in the emitter and the absorber, respectively. $D_\infty = \int_{\partial V} dr^2 \frac{1}{2}\text{Re}[\mathbf{E_n}(r) \times \mathbf{H_n^\dagger}(r)]$ has the form of the mode energy loss due to far-field radiation. $F$ is a factor attributed to the non-Hermitian imperfection of the lossy resonant mode expansion, which equals $F = \left|\int_V dr^3 \left\{\text{Re}\left[\frac{\partial \omega \epsilon}{\partial \omega}(Re[\omega_n])\right] |\mathbf{E_n}|^2 + \mu_0 |\mathbf{H_n}|^2\right\}\right| / |N_{nn}|$.

Next, we investigate the far-field thermal radiation of $V_E$ in Eq. (5) with the substitution of Eqs. (7) and (8). For the frequency $\omega$ close to the resonant frequency Re[$\omega$] of the mode, the first term on the RHS of Eq. (5) becomes

$$4\omega\mu_0 \text{Tr}\left[\int_{V_E} dr^3 \sigma_E \text{Im}\left[\bar{\mathbf{G}}_{\omega,\mathbf{r},\mathbf{r}}\right]\right] \approx 4 \text{Im}\left[\frac{1}{(\omega_n - \omega)}\right] \text{Re}\left[\frac{\int_{V_E} dr^3 \sigma_E(\omega) \mathbf{E_n}(r)^T \mathbf{E_n}(r)}{N_{nn}}\right], \quad (13)$$
$$= L(\omega)\psi_1(\omega)$$

where the peak value $\Psi_1 = \psi_1(\text{Re}[\omega_n]) = 4\left(\frac{D_E}{D_E+D_A+D_\infty}F\right)P$, and the factor $P$ is defined as

$$P = \text{Re}\left[\frac{\int_{V_E} dr^3 \sigma_E \mathbf{E_n}(r)^2}{N_{nn}}\right] / \frac{\int_{V_E} dr^3 \sigma_E |\mathbf{E_n}(r)|^2}{|N_{nn}|}, \quad (14)$$

where $\mathbf{E_n}^2 = \mathbf{E_n}^T \cdot \mathbf{E_n}$. $P$ is the other imperfection factor due to the non-Hermitian fact as compared to the factor $F$. Mathematically, it has $P \leq 1$. Since the second and the third terms in Eq. (5) have the same form with Eq. (4), the spectral thermal radiation of $V_E$ to far-field can be derived from Eqs. (12) and (13) as $\phi_\infty(\omega) = \frac{\Theta}{2\pi}L(\omega)\psi_\infty(\omega)$, where the peak value $\Psi_\infty = \psi_\infty(\text{Re}[\omega_n])$ equals

$$\Psi_\infty = 4\left(\frac{D_E}{D_E+D_A+D_\infty}F\right)\left(P - \frac{D_E+D_A}{D_E+D_A+D_\infty}F\right). \quad (15)$$

Hence, Eqs. (12) and (15) represent a new theoretical framework to understand and control the thermal radiation from optical resonators in both near- and far-fields. To interpret the weights of the mode energy losses into all possible sources, we define the *fractional mode loss* of the emitter, the near-field absorber, and the far-field as $\eta_E = \frac{D_E}{D_E+D_A+D_\infty}F$, $\eta_A = \frac{D_A}{D_E+D_A+D_\infty}F$ and $\eta_\infty = P - \eta_E - \eta_A$, respectively.

Although the values of $D_\infty$ and $F$ depend on the choices of the enclosure surface $\partial V$ and the volume $V$, $\eta_E$, $\eta_A$ and $\eta_\infty$ all have fixed values because $\frac{F}{D_E+D_A+D_\infty} = \left|\frac{4}{\text{Im}[\omega_n]N_{nn}}\right|$ according to Eq. (11). As a result, Eqs. (12) and (15) become

$$\begin{aligned}\Psi_A &= 4\eta_E\eta_A \\ \Psi_\infty &= 4\eta_E\eta_\infty\end{aligned}, \quad (16)$$

given that both $\Psi_A$ and $\Psi_\infty$ are positive, and $\eta_E + \eta_A + \eta_\infty = P$, the maxima of both $\Psi_A$ and $\Psi_\infty$ are equal to $P^2$, when $\eta_E = \eta_A = \frac{P}{2}$ and $\eta_\infty = 0$ for near-field thermal radiation, and $\eta_E = \eta_\infty = \frac{P}{2}$, $\eta_A = 0$ for far-field thermal radiation.

Because of the modulation of a single resonant mode, both near-field and far-field spectral thermal energy fluxes follow the Lorentzian line shape. To maximize thermal emission, Eqs. (12) and (15) demonstrate that the fractional mode losses must be matched in order to achieve maximized thermal radiation, i.e. $\eta_E = \eta_A = \frac{P}{2}$ for near-field emission, and $\eta_E = \eta_\infty = \frac{P}{2}$ for far-field emission. In our theory, the peak or maximum value of both near-field and far-field spectral thermal emission equals $\frac{\Theta}{2\pi}P^2$, which is the key distinction from coupled-mode theory, i.e. the peak value equals $\frac{\Theta}{2\pi}$ [11,12]. To maximize $P$ for reaching the limit of the peak value, Eq. (14) suggests that optical resonator thermal emitters need to be designed to have electrical quasi-static resonant modes, i.e. the electric field oscillates in phase, or $\mathbf{E_n}(r)$ near the optical resonator has a real value and it can be expressed by a voltage potential as $\mathbf{E_n}(r) = -\nabla\Phi(r)$ [10]. In this case, we prove $P \approx 1$ and $F \approx 1$ because (i) by choosing V to only enclose the near-field components of the resonant mode, the portion of the resonant mode outside $V$ behaves as the propagating waves and therefore $\left(\frac{\partial\omega\epsilon}{\partial\omega}\right)\mathbf{E_n^2} = \epsilon\mathbf{E_n^2} \approx \mu\mathbf{H_n^2}$, which results in $N_{nn} \approx \int_V dr^3 \left(\frac{\partial\omega\epsilon}{\partial\omega}\right)\mathbf{E_n^2} - \mu\mathbf{H_n^2}$; (ii) the portion of the resonant mode inside $V$ behaves to be quasi-static, where $\mathbf{H_n} = \frac{1}{i\omega\mu_0}\nabla\times\mathbf{E_n} \approx 0$, and thus $N_{nn} \approx \int_V dr^3 \left(\frac{\partial\omega\epsilon}{\partial\omega}\right)\mathbf{E_n^2}$; (iii) Since the quasi-static electric field has a real value, $\int_{V_E}\sigma_E\mathbf{E_n^2} \approx$ $\int_{V_E} \sigma_E |\mathbf{E_n}|^2$ and $N_{nn} \approx \int_V dr^3 \left(\frac{\partial \omega \epsilon}{\partial \omega}\right) |\mathbf{E_n^2}| \approx |N_{nn}|$ based on the fact $\frac{\partial \omega \epsilon}{\partial \omega} > 0$. Likewise, $F \approx \left|\frac{\int_V dr^3 \text{Re}\left[\frac{\partial \omega \epsilon}{\partial \omega}\right] |\mathbf{E_n}|^2}{\int_V dr^3 \left(\frac{\partial \omega \epsilon}{\partial \omega}\right) \mathbf{E_n^2}}\right| \approx 1$. As a result, in the quasi-static condition, our formalism agrees with coupled mode theory, but has the direct electromagnetic definition for each mode energy loss.

In order to validate our formalism, we investigate the far-field and near-field thermal emission spectra of the optical resonators made from metal nanorods, where the theoretical predictions by Eqs. (12) and (15) are compared with the direct simulations by the Fluctuating-Surface Current (FSC) method [6,23]. The field profile and the eigen-frequency of the lossy resonant modes are numerically calculated by the fully vectorial finite-element method [20], and the fractional mode losses $\eta_E$, $\eta_A$ and $\eta_\infty$ can then be evaluated accordingly. Consider the far-field thermal radiation of a gold nanorod in the spectral range corresponding to its fundamental resonant modes (Figs. 2(a) and (b)), which is indeed the Fabry-Perot resonance of the $TM_0$ waveguiding mode. The length of the nanorods is kept as $L = 2.5\mu m$, and the material is chosen to have a Drude model with $\omega_p = 1.37 \times 10^{16} \text{rad} \cdot \text{s}^{-1}$ and $\gamma = 5.32 \times 10^{13} \text{rad} \cdot \text{s}^{-1}$. The predicted and the simulated emission spectra of the nanorods with different diameters $D$ are plotted in Fig. 2(c). The good agreement between our predictions and the direct simulations convincingly verifies our formalism in the far-field case. Figure 2(d) further investigates the values of $\eta_E$ and $\eta_\infty$ at different diameters $D$. The monotonic increase of $\eta_E$ is attributed to the fact that the dissipative loss of the $TM_0$ waveguiding mode increases as the decrease of the waveguide lateral size $D$ [24]. At $D \sim 40$nm, the radiative and the resistive mode losses are matched, and meanwhile $P \to 1$, i.e. $\eta_E = \eta_\infty \sim 0.5$. Consequently, the thermal emission peak reaches to the theoretical limit $\frac{\Theta}{2\pi}$. According to our aforementioned proof, the reason of $P \to 1$ at $D \sim 40$nm is that the electric field of the resonant mode inside the metal nanorod is approximately quasi-static with the uniform phase, as shown in Fig. 2(e). However, for a thick nanorod, e.g. $D = 200$nm, the phase of the electric field varies significantly in the cross-sections and consequently $P = 0.9 < 1$.

Next, we investigate the near-field radiative transfer between two gold nanorods separated by a distance of 50 nm. By calculating the fundamental resonant mode in Figs. 3(a) and (b), the energy transfer spectra at different diameter $D$ are predicted in Fig. 3(c), which agree well with the direct simulations results and therefore our formalism in the near-field case is also verified. The "bump" around the first peak of red curve in Fig. 3(c) is attributed to a degenerate "symmetric" mode (See Supplemental Materials). Figure 3(d) plots the fractional mode losses at different $D$, where $\eta_E = \eta_A$ is because the emitter and the absorber are exactly the same, and $\eta_E = \eta_A \to 0.5$ for $D \to 0$ is due to the same reason as that of the single nanorod cases, i.e. the dissipative loss of the $TM_0$ waveguiding mode is monotonically increasing as the decrease of the lateral size. Overall, our theory can well predict both far-field and near-field thermal radiation spectra under the assumption of *single predominant resonance mode*. In general, this assumption is valid if (i) the resonance modes of the structure have distinct resonance frequencies; and (ii) the resonance mode accounts for the major energy decay channel of thermal radiation, which implies that a significant portion of the mode field profile is concentrated around the emitter [15], as shown in Figs. 2(a) and 3(a) from the FEM calculations. In addition, the small deviation between our theory and the direct simulations in Figs. 2(c) and 3(c) is mainly because the non-resonance contribution to thermal radiation is ignored in our theory.

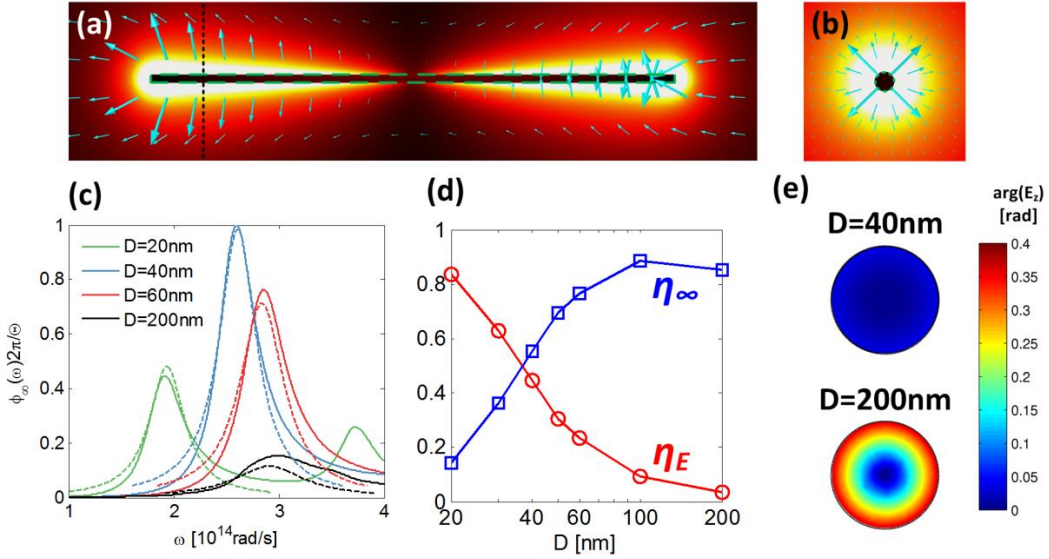

FIG. 2: (color online) (a) and (b) Electric field profiles of the fundamental resonant mode of the gold nanorod with $D = 40$nm and $L = 2.5$μm. The color profile and the arrows indicate the field intensity and polarization, respectively. (c) Spectral energy flux of thermal emission evaluated based on our theory (dash curves) and direct calculation (solid curves). (d) Fractional mode losses for the cases with different $D$. (e) Electric field phase profile of the resonant mode in the cross-section of the gold nanorod.

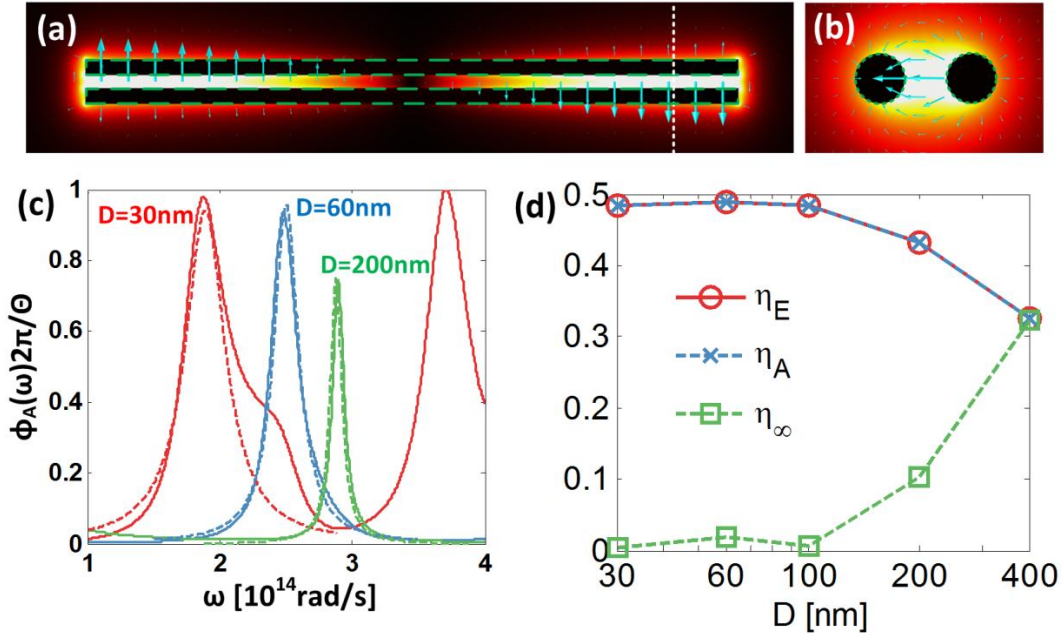

**FIG. 3: (color online) (a) and (b) Electric field profiles of the fundamental resonant mode of two aligned gold nanorods with $D = 60\text{nm}$, $L = 2.5\text{μm}$ and a 50nm gap. Color profile and the arrows indicate the field intensity and polarization, respectively. (c) Spectral energy flux of thermal emission evaluated based on our theory (dash curves) and direct calculation (solid curves). (d) Fractional mode losses for the cases with different $D$.**

In conclusion, we develop a general formalism from the fluctuational electrodynamics and quasi-normal mode theory to elucidate the underlying physics of the far-field and near-field thermal radiation from the optical resonators made by lossy and dispersive material. Because of the modulation from the resonant mode, the thermal emission power density spectrum of the optical resonators is narrow-band and follows the Lorentzian line shape with the peak at the resonant frequency of the mode. To maximize the thermal emission, our formalism demonstrates that not only the mode losses to the emitter and the absorber (or far-field background) require to be matched, but the resonant mode also needs to be electrically quasi-static, i.e. the electric field of the resonant mode oscillates in phase. In addition, we validate our formalism by investigating the far-field and near-field thermal emission from metal nanorods. Our formalism can therefore serve as a general theoretical framework to design the narrow-band thermal emission of arbitrary resonant structures.

The authors acknowledge funding support from US National Science Foundation (CBET-1253692).